\documentstyle[aps,prl,multicol]{revtex}
\begin{document}
\draft
\title{Superconducting correlations in 
 metallic nanoparticles:
exact solution of the BCS model 
by the algebraic Bethe ansatz}

\author { 
Huan-Qiang Zhou\cite{email0}, Jon Links, Ross
H. McKenzie, and Mark D. Gould}

\address{Centre for Mathematical Physics, The University of Queensland,
		     Brisbane, Qld 4072, Australia}

\maketitle

\vspace{10pt}

\begin{abstract}
Superconducting pairing of electrons in nanoscale
metallic particles with discrete energy levels
and a fixed number of electrons is described 
by the reduced BCS model Hamiltonian.
We show that this model is integrable by
the algebraic Bethe ansatz.
The eigenstates, spectrum, conserved operators,
integrals of motion, and norms of wave functions
are obtained.
Furthermore, the quantum inverse problem
is solved, meaning that form factors and correlation
functions can be explicitly evaluated.
Closed form expressions are given for the form factors
and correlation functions that describe superconducting pairing.
\end{abstract}

\pacs{PACS numbers: 71.24+q, 74.20Fg}



\def\a{\alpha}
\def\b{\beta}
\def\d{\dagger}
\def\e{\epsilon}
\def\g{\gamma}
\def\k{\kappa}
\def\l{\lambda}
\def\o{\omega}
\def\t{\tilde{\tau}}
\def\s{S}
\def\D{\Delta}
\def\L{\Lambda}
\def\T{{\cal T}}
\def\TT{{\tilde{\cal T}}}

\def\beq{\begin{equation}}
\def\eeq{\end{equation}}
\def\bea{\begin{eqnarray}}
\def\eea{\end{eqnarray}}
\def\ba{\begin{array}}
\def\ea{\end{array}}
\def\no{\nonumber}
\def\le{\langle}
\def\re{\rangle}
\def\lt{\left}
\def\rt{\right}

\newcommand{\reff}[1]{eq.~(\ref{#1})}

\begin{multicols}{2}

Due to recent advances in nanotechnology it
has become possible to fabricate and characterise
individual metallic grains with dimensions as
small as a few nanometers~\cite {dr01}.
They are sufficiently small that the spacing,
$d$, of the discrete energy levels can be
determined.
A particularly interesting
question concerns whether superconductivity can 
occur in a grain with $d$  comparable to
$\Delta$, the energy gap in a bulk system.
If $d \ll \Delta$, the superconducting correlations
are well-described by a mean-field solution
to the reduced pairing Hamiltonian
(equation (\ref{BCSH}) below)
due to Bardeen, Cooper, and Schrieffer (BCS)
 in the grand canonical
ensemble with a variable number of electrons.
However, if $d \sim \Delta$ 
recent numerical calculations have shown
that when the number of electrons is fixed
(as in the canonical ensemble) 
the superconducting
fluctuations become large and approximate 
treatments become unreliable\cite{dr01,appro}.
Thus, exact calculations of physical
quantities are highly desirable.
It has only recently been appreciated that the
exact eigenstates 
and spectrum of the BCS 
model were found  in the
1960's by Richardson, in the context of
nuclear physics~\cite{dr01,rs64}.
The model has subsequently been found to
have a rich mathematical structure:
it is integrable (i.e., has  a complete set of conserved
operators) \cite{crs97}, has a connection to conformal field
theory \cite{s00}, and is related to Gaudin's inhomogeneous
spin-1/2  models \cite{alo01,ao01,aff01,s89}.

In this Letter we show how the BCS model
can be solved using the algebraic Bethe ansatz (ABA) method.
This result can be deduced from the observation that the conserved
operators
obtained in \cite{crs97} were also obtained in \cite{s89} via the ABA, 
but in another context. However, the approach we adopt here is
slightly different
from \cite{s89}, which 
facilitates the solution of the quantum inverse problem
\cite{kib,kmt98,mtgk} to explicitly evaluate form factors
(i.e., one point functions) and correlation
functions. 
This completes the agenda recently set out
by Amico, Falci, and Fazio \cite{aff01}.
We also readily obtain known results
for eigenstates, the spectrum, and conserved operators.
  Our treatment is also applicable to 
superconductivity in fermionic atom traps 
  \cite{stoof,truscott} and can also
be extended to a solvable model
for condensate fragmentation in boson systems \cite{ds01}.

The Hamiltonian for the
reduced BCS model consists of a kinetic energy
term and an interaction term which
describes the attraction between electrons in time reversed states,
\bea
H_{BCS}&=& \sum ^{\Omega}_{\stackrel {j=1}{\sigma=+, -}}
\e_{j}  c^\d_{j\sigma}c_{j\sigma}
 -g\sum ^{\Omega}_{j,j'=1} c^\d_{j+}c^\d_{j-}
 c_{j'-}c_{j'+},\label{BCSH}
\eea
where $j=1, \cdots, \Omega$ labels a shell of doubly degenerate single
particle energy levels with energies $\e_j$ and $c_{j\sigma}$ the
annihilation operators; $\sigma= +, -$ labels the degenerate
time reversed states;
 $g$ denotes the BCS pairing coupling constant.
Using the pseudo-spin realization of electron pairs:
$\s^z_j= (c^\d_{j+}c_{j+}+
 c^\d_{j-}c_{j-}-1)/2,
\s^+_j= c^\d_{j+}c^\d_{j-}$ and
$ \s^-_j= c_{j-}c_{j+}$, the BCS Hamiltonian
(\ref{BCSH}) becomes (up to a constant term) 
\beq
H_{spin}= \sum ^{\Omega}_{j=1}
2\e_{j} \s^z_j-\frac {g}{2}
\sum ^{\Omega}_{j,k=1} (\s^+_j \s^-_k + \s^+_k \s^-_j). 
\label{spinH} \eeq

{\it The $R$ matrix.}        
 An essential ingredient of the ABA, which follows from the
Quantum Inverse Scattering Method (QISM), is the
construction of the $R$
matrix solving the quantum Yang-Baxter equation,
\bea 
&& R _{12} (u_1-u_2)  R _{13} (u_1-u_3)  R _{23} (u_2-u_3) \no \\
&& ~~~= R _{23} (u_2-u_3)  R _{13}(u_1-u_3)  R _{12} (u_1-u_2)
\no \eea 
where the $u_j$ are spectral parameters.
Here $R_{jk}$ denotes the matrix on $V \otimes V\otimes V$ 
(where $V$ is the two-dimensional Hilbert space
on which the pseudo-spin operators act) acting on the
$j$-th and $k$-th spaces and as an identity on the remaining space.
The $R$ matrix may be viewed as the structural constants for the
Yang-Baxter algebra generated by the monodromy matrix $T(u)$,
\beq
R_{12}(u_1-u_2) T_1(u_1) T_2(u_2)=
T_2(u_2) T_1(u_1)R_{12}(u_1-u_2). \label{YBA}
\eeq
There are two kinds of realisations of the Yang-Baxter algebra which are
relevant to our construction. One is operator-valued given by the $R$ matrix
$R_{0j}(u)$ and  the other is a c-number representation $G$ 
which does
not depend on the spectral parameter $u$. In the latter case, we have
$[R(u),G\otimes G]=0$. The comultiplication behind the Yang-Baxter
algebra allows us to construct a representation of  the monodromy matrix
through
$$
T_{0}(u) = G_0 R_{0\Omega}(u-\e_{\Omega})\cdots
G_0 R_{01}(u-\e_1).
$$
Defining the transfer matrix via $t(u) \equiv tr_0 T_0(u)$ it follows
that  $[t(u),t(v)]=0$ for all values of the parameters $u,\,v$. 
If the $R$ matrix possesses the
regularity property 
$ R_{jk}(0)= P_{jk}$ with $P$ being the permutation operator,
then it is easily verified  that
\bea
t(\e_j)&=& G_jR_{jj-1}(\e_j-\e_{j-1}) \cdots 
 G_j R_{j1}(\e_j-\e_1)\no\\   
&& \times  G_jR_{j\Omega}(\e_j-\e_{\Omega}) \cdots 
 G_j R_{jj+1}(\e_j-\e_{j+1}) G_j.
\label{t1}
\no \eea

Let us now assume that the $R$ matrix is quasi-classical, i.e.,
it admits a series expansion $R(u)=I+\eta r(u)+\cdots$, for an 
appropriate parameter $\eta$. If we can also choose $G$ such that 
$G=1+\eta \Gamma+ \cdots$, then the expansion of $t(\e_j)$ in terms of 
$\eta$ takes the form,
\beq
t(\e_j)= I+\eta \tau_j +\cdots.
\label{t2}
\eeq
An immediate consequence from the commutativity of the transfer matrics 
is $[\tau_j,\tau_k]=0$. Therefore an integrable model
is obtained by taking the set $\{\tau_i\}$ as the conserved operators 
and a Hamiltonian
given as a function of the $\tau_j$.

We apply the procedure described above to the $su(2)$ 
invariant $R$ matrix $R(u) = b(u) I + c(u) P$,
with entries that are rational functions:
$b(u)=u/(u+\eta)$ and $c(u)={\eta}/(u+\eta)$.
Note that the regularity property $R(0)=P$ is present. 
For this case, we can choose  $G$ as any element of the $su(2)$
algebra.
We claim that the BCS model corresponds to
the special choice
\beq
 G_j=\exp(-2 \eta \s^z_j/g\Omega).
\eeq
This can be viewed as a generalized inhomogeneous
six-vertex model.
Expanding this and the $R$ matrix to first order in $\eta$
and substituting in (\ref{t1}) we find from (\ref{t2}) that
$$
 \tau_j= -\frac{2}{g} \s^z_j + 2 \sum ^{\Omega}_{k \neq j}
\frac {{\bf \s}_j \cdot {\bf \s}_k}{(\e_j-\e_k)}
$$
where we have discarded a constant term. 
These operators are the isotropic Gaudin Hamiltonians in a
non-uniform magnetic field ~\cite{s89}.
Their relevance to the spin realization of
the BCS model (\ref{spinH}) is that the latter is expressible  
(up to a constant) as 
$$H_{spin}= -g \sum_{j=1}^{\Omega}(\e_{j}-g/2)\tau_j +\frac {g^3}{4}
\sum_{j,k=1}^\Omega \tau_j \tau_k.
$$    
Although the above expressions for the consereved operators only applies
to  the case when all $\e_j$'s are distinct, our construction can be
adapted to accommodate the cases when some of $\e_j$'s are the same.

{\it Algebraic Bethe ansatz.}
In the ABA, the integrals of motion are obtained by
finding the eigenfunctions of the
transfer matrix which is given by the trace of the 
monodromy matrix. The monodromy matrix is written in the form
$$
T(u) = \left ( \begin {array} {cc}
A(u)&B(u)\\
C(u)&D(u)
\end {array} \right ),
$$
which is the quantum equivalent of the scattering 
coefficients of the classical inverse scattering
problem.
Then from the Yang-Baxter algebra (\ref{YBA}), we may derive the fundamental
commutation relations (FCR) between the entries of the monodromy matrix.
Choosing the state $|0\rangle =\otimes ^\Omega _{j=1} 
|\uparrow \rangle_j $
as the pseudovacuum, then we have the pseudovacuum eigenvalues $a(u)$ and $d(u)$
of $A(u)$ and $D(u)$:
$a(u)=  \exp (-\eta/g),
d(u)= \exp (\eta/g) \prod _j b(u-\e_j)$.
Following the standard procedure \cite{kib,f80}, we choose the Bethe
state 
\beq
\Psi (v_1,\cdots, v_N) = \prod ^N_{\a =1} B(v_\a) |0 \rangle.  
\eeq
Then we may derive the off-shell Bethe ansatz equations 
using the FCR following \cite{f80,b93}, which, in the quasiclassical limit,
takes the form
\beq
\frac {1}{2} \tau_j \psi = \l_j \psi - 
\sum ^N_{\a=1} \frac {f_\a \s^-_j}{\e_j-v_\a}
\psi '_\a, \label{qc-osbae} 
\eeq
where
\bea
\l_j &=& -\frac {1}{2g} -\frac {1}{2} \sum _\a \frac {1}{\e_j - v_\a} +
\frac {1}{4}\sum _{i \neq j} \frac {1}{\e_j - \e_i}, \no\\
f_\a &=& \frac {1}{g} +  \sum _{\b \neq \a} \frac {1}{v_a -v_\b}
-\frac {1}{2} \sum _j \frac {1}{v_\a -\e_j}, \no\\
\psi &\equiv&|v_1,\cdots,v_N \rangle =
\prod ^N_{\a=1} \sum ^\Omega _{j=1} \frac {\s^-_j}{v_\a -\e_j}
|0 \rangle.
\no \eea
In (\ref{qc-osbae}) we defined 
$\psi '_\a$ by 
$$\psi = \sum ^\Omega
_{j=1} \frac {\s ^-_j}{v_\a-\e_j} \psi '_\a. $$ 
Imposing $f_\a =0$, one immediately sees that $\psi$ becomes the
eigenvector of the conserved operator $\tau _j$ with
$\l_j$ as the eigenvalue. The constraint $f_\a=0$ is then
equivalent to Richardson's equations \cite{rs64},
\beq
\frac {2}{g}+ \sum ^N_{\b \neq \a} \frac {2}{v_\a-v_\b}=
\sum ^\Omega _{j=1} \frac {1}{v_\a-\e_j}. \label{bae}
\eeq
Here $L-N$ may be interpreted as the number of
time-reversed pairs of electrons.
The energy eigenvalue of the Hamiltonian (\ref{spinH}) is
\beq
 E_{spin}
=\sum_{j=1}^\Omega \e_j -2\sum_{\a=1}^N  v_\a +g(2N-\Omega).
\eeq

{\it Scalar products and norms.}
Directly evaluating
the norms of Bethe wave functions
can be tedious, if not impossible.
However, using the QISM
they can be represented as determinants ~\cite{kib,efik}.
Since this representation only depends 
on the $R$ matrix, the derivation
presented previously for different models
can be readily applied to our (generalized)
inhomogeneous six-vertex model.
 In the QISM
construction, the determinant representation for scalar products
$$\langle 0| \prod ^N_{\b=1} C(w_\b) \prod^N_{\a=1} B(v_\a) |0
\rangle $$
play a crucial role; especially, when one of the sets of
parameters, for example $\{v_a\}$, is a solution of the Bethe equations
\cite{kib,kmt98,sla89}.
 In the quasiclassical limit, the leading
term of the scalar product for the inhomogeneous six-vertex model
gives rise to the scalar product 
\bea
&& \langle w_1,\cdots,w_N|v_1,\cdots,v_N\rangle = \no\\
&&~~~\frac {\prod ^N_{\b=1} \prod ^N_{\stackrel {\a=1}{\a \neq \b}}
(v_\b-w_\a)}
{\prod _{\b <\a} (w_\b -w_\a) \prod _{\a <\b} (v_\b -v_\a)}
{\rm det}_N J(\{ v_\a \}, \{ w_\b \}),\label{sp-21}
\eea
where the matrix elements of $J$ are given by
\bea
J_{ab} &=& \frac {v_b -w_b}{v_a -w_b} 
\left ( \sum ^\Omega _{j=1} \frac {1}
{ (v_a -\e _j)(w_b -\e _j)} \right. \no\\
&& \left. -2\sum _{\a \neq a} \frac {1}{(v_a
-v_\a)(w_b -v_\a)} \right ).
\label{sp-22}
\eea
Here $\{ v_\a \}$ are a solution to Richardson's equations (\ref{bae}),
whereas $\{ w_\b \}$ are arbitrary parameters.
 Richardson's expression \cite{r65} for the square of the
norm of the Bethe state follows from (\ref{sp-21}) and (\ref{sp-22})
by taking the limit $w_\a \rightarrow v_\a$.

{\it Solution of  the quantum inverse problem.} 
In order to calculate the form factors and correlation functions, we
need to solve this problem for the
generalized inhomogeneous six-vertex model. This then allows the
reconstruction of local quantum spin operators in terms of the quantum
monodromy matrix. 
A general procedure for doing
this has recently been presented \cite{kmt98,mtgk} for the so-called
{\it fundamental} models. In our case, where the model is not
fundamental, we find
\bea
S^-_i &=& \prod^{i-1}_{\a=1} t(\e_\a) K^{-i+1}B(\e_i) K^{i-1}
\prod ^i_{\a=1} t^{-1}(\e_\a),\no\\ 
S^+_i &=& \prod^{i-1}_{\a=1} t(\e_\a) K^{-i+1}C(\e_i) K^{i-1}
\prod ^i_{\a=1} t^{-1}(\e_\a),\no\\
S^z_i &=& \prod^{i-1}_{\a=1} t(\e_\a) K^{-i+1} 
\frac {(A(\e_i)-D(\e_i))}{2} K^{i-1}
\prod ^i_{\a=1} t^{-1}(\e_\a), \no 
\eea
with $ K \equiv \prod ^\Omega _{j=1} G_j = \exp (-2\eta \sum
^\Omega_{j=1} S^z_j/g\Omega)$. The above
construction is one of our main results.
The appearance of the powers of $K$
arises from the c-number matrix realisation of the Yang-Baxter algebra
$G$ which is peculiar to our construction.  
Following \cite{kmt98}, one can
obtain the representation of the correlation functions in terms of
pseudovacuum eigenvalues $a(u)$ and $d(u)$. 

{\it Form factors.}
For the BCS model
the pair correlator
\bea
C^2_m \equiv \langle c^\d_{m+} c_{m+} c^\d_{m-} c_{m-}\rangle 
- \langle c^\d_{m+} c_{m+} \rangle
 \langle c^\d_{m-} c_{m-} \rangle \label {pc}
\eea
is of particular interest \cite{dr01,braun}.
(We use the notation that
$\langle \chi \rangle \equiv 
\langle v_1,\cdots v_N| \chi |v_1,\cdots,v_N \rangle /
\langle v_1,\cdots v_N|v_1,\cdots,v_N \rangle $ for any operator
$\chi$).
$C^2_m$
can be interpreted as the probability
enhancement of finding a pair of electrons 
in level $m$, instead of two uncorrelated electrons.
(It is zero for $g=0$). 
In the pseudo-spin representation
$C^2_m = {\langle S^-_m S^+_m \rangle
\langle S^+_m S^-_m \rangle} = { 1/4-\langle S^z_m
\rangle^2}$.
In general, form factors such as
\bea
F^z(m, \{w_\b \},\{ v_\a \}) &\equiv& \langle 0| \prod ^N_{\b=1} 
C(w_\b)  S^z_m \prod^N_{\a=1} B(v_\a) |0 \rangle \no 
\eea
can be calculated for the generalized inhomogeneous six-vertex model. 
In the quasiclassical limit, they reduce
to the form factors of the BCS model,
\bea
&& \langle w_1,\cdots,w_{N+1}|S^-_m|v_1,\cdots,v_N\rangle = \no\\
&& \langle v_1,\cdots,v_N|S^+_m|w_1,\cdots,w_{N+1}\rangle =\no \\
&& \frac {\prod ^{N+1}_{\b=1} (w_\b - \e_m)} 
{\prod ^N_{\a=1} (v_\a - \e_m)} 
\frac { {\rm det}_{N+1} \T (m, \{ w_\b \}, \{ v_\a \})}
{\prod _{\b > \a} (w_\b -w_\a) \prod _{\b <\a} (v_\b -v_\a)},\no\\
&& \langle w_1,\cdots,w_N|S^z_m|v_1,\cdots,v_N\rangle = 
\prod ^N_{\a=1} \frac {(w_\a - \e_m)} 
{(v_\a - \e_m)}\no\\ 
&&~~~ \times\frac { {\rm det}_N \left (\frac {1}{2} \TT(\{ w_\b \}, \{ v_\a \}) 
- Q (m, \{ w_\b \}, \{ v_\a \}) \right )} 
{\prod _{\b > \a} (w_\b -w_\a) \prod _{\b <\a} (v_\b -v_\a)},
\no \eea

with the matrix elements of $\T$ given by
\bea
\T_{ab}(m) =&& 
\prod ^{N+1}_{\stackrel {\a=1}{\a \neq a}} (w_\a - v_b)
\left ( \sum ^\Omega _{\j=1} \frac {1}
{ (v_b -\e _j)(w_a -\e _j)} \right. \no\\ 
&& \left. -2\sum _{\a \neq a} \frac {1}{(v_b-
w_\a)(w_a -w_\a)} \right ),
~~~b < N+1, \no\\
\T_{aN+1}(m)  && =  \frac {1}{(w_a -\e_m)^2}, \ \
Q_{ab}(m) = \frac {\prod _{\a \neq b} (v_\a-v_b)} {(w_a-\e_m)^2}.
\no \eea
Above, $\TT$ is the $N\times N$ matrix 
obtained from $\T$ by deleting the last row and column and replacing $N+1$ by
$N$ in the matrix elements. Here we assume that both $\{ v_\a \}$ and
$\{ w_b \}$ are solutions to Richardson's Bethe equations
(\ref{bae}). However, the results are still valid for $S^\pm_m$ if only 
$\{ w_b \}$ satisfy the Bethe equations.

{\it Correlation functions.}
We find that the correlation functions of 
the BCS model
take the same form as the underlying $su(2)$ spin 1/2 Gaudin model,
with the parameters $v_j$ satisfying Richardson's Bethe ansatz equations
(\ref{bae}) instead of
Gaudin's ones. Here we present explicitly the two-point correlation function
\bea
&& \langle w_1,\cdots,w_N|S^-_mS^+_n|v_1,\cdots,v_N\rangle = \no\\
&& \sum _{\a=1}^N \frac {1}{v_\a -\e_n}
\langle w_1,\cdots,w_N|S^-_m|v_1,\cdots,\hat v _\a, \cdots, v_N\rangle -\no\\
&& \sum _{\a \neq \b} \frac {1}{(v_\a -\e_n)(v_\b -\e_n)} \times \no\\
&&\langle w_1,\cdots,w_N|S^-_mS^-_n|v_1,\cdots,\hat v _\a,
\cdots, \hat v _\b, \cdots, v_N\rangle.
\eea
Here the hat denotes that the corresponding parameter is not present in
the set. Since $\{ w_a \}$ is a solution of the Bethe equations, 
$\langle w_1,\cdots,w_N|S^-_m|v_1,\cdots,\hat v _\a, \cdots, v_N\rangle $
is the form factor given before, while
\bea
&& \langle w_1,\cdots,w_N|S^-_mS^-_n|v_1,\cdots,v _{N-2} \rangle =\no\\
&& \frac {\prod ^N_{\b=1} (w_\b - \e_m)(w_\b-\e_n)} 
{\prod ^{N-2}_{\a=1} (v_\a - \e_m)(v_\a -\e_n)} 
\frac { {\rm det}_N \T (m,n, \{ w_\b \}, \{ v_\a \})}
{\prod _{\b > \a} (w_\b -w_\a) \prod _{\b <\a} (v_\b -v_\a)},\no\\
\label {2ff} \eea
with
\bea
\T_{ab}(m,n) =&& 
\prod ^N_{\stackrel {\a=1}{\a \neq a}} (w_\a - v_b)
\left ( \sum ^\Omega _{\j=1} \frac {1}
{ (v_b -\e _j)(w_a -\e _j)} \right. \no\\ 
&& \left. -2\sum _{\a \neq a} \frac {1}{(v_b-
w_\a)(w_a -w_\a)} \right ),
~~~b < N-1, \no\\
\T_{aN-1}(m,n)  && =  \frac {2w_a-\e_m-\e_n}{[(w_a -\e_m)(w_a-\e_n)]^2},\no\\ 
\T_{aN}(m,n)  && =  \frac {1}{(w_a -\e_m)^2}, 
\no \eea
In (\ref{2ff}) $m \neq n$ is assumed, with the convention that it is
zero when $m=n$. The above results constitute the building blocks of the
Penrose-Onsager-Yang off-diagonal longe-range order (ODLRO) parameter
$\Delta_{OD}$ \cite{od},
\beq
\Delta_{OD} \equiv \frac {1}{\Omega} \sum _{mn} \langle S^+_nS^-_m \rangle.
\eeq
The small grain behavior of this parameter and its connection with the
pair correlator (\ref{pc}) was recently  discussed  in
\cite {ttc}.

{\it Further applications.}
Our work  is also relevant to
proposals to observe BCS superconductivity
in gases of fermionic atoms such as spin-polarised
$^6$Li \cite{stoof}. Quantum degeneracy of $^6$Li at
temperatures of about 240 nK has recently
been observed in an atom trap with frequencies,
$\omega \sim $ 1 kHz \cite{truscott}, corresponding
to an energy level spacing of the order of
$10^{-12}$ eV.
The estimated BCS transition temperature is
of the order of 20 nK \cite{stoof}, corresponding
to an energy gap of the order of 
$4 \times 10^{-12}$ eV.
Hence, these systems are in a regime where
the physics considered here will be important.

Dukelsky and Schuck \cite{ds01} recently introduced
a solvable model for condensate fragmentation in    finite
boson systems.
The model they solved follows from
the construction used above
when the Yang-Baxter algebra is realized in terms of
the generators of the Lie algebra $su(1,1)$. The model also provides
a new mechanism for the enhancement of $sd$ dominance in interacting
boson models in the context of nuclear physics \cite{dp01}.

This work was supported by the Australian Research Council.
 \end{multicols}

\end{document}